\documentstyle[preprint,aps,version2]{revtex}   % von JB
\begin{document}
\begin{title}
Probing the  spin polarization in ferromagnets
\end{title}
\date{ }
\author{ Jamal Berakdar\thanks{e-mail:jber@mpi-halle.de}}
\begin{instit}  Max-Planck-Institut f\"ur
Mikrostrukturphysik, Weinberg 2, 06120 Halle, Germany
\end{instit}
\begin{abstract}
The emission of correlated electrons
from an itinerant ferromagnet following the impact of a polarized electron beam
 is analyzed in terms of irreducible tensorial parameters that can be 
measured. Under favorable  conditions, specified in this work, these
parameters are related to the spin polarization in the ferromagnet.
The formal results are illustrated by numerical studies of the
polarized electron pair emission from a Fe(110) surface and 
a novel technique for the investigation of
 magnetic properties of  ferromagnets is suggested.
\end{abstract}
%
%\begin{multicols}{2}
%\begin{document}
 The  electronic and magnetic properties of low dimensional
systems with long-range magnetic order, such as  ultrathin
ferromagnetic films and magnetic surfaces are currently
under intensive investigations \cite{hubert}. This is due to the 
fundamental and
technological importance of such materials. 
Magnetic systems
with reduced symmetry  can be explored by a variety of techniques \cite{hubert}.
Low-energy spin-polarized  
  electron spectroscopy is particularly suitable  
as  the penetration depth  is on the order of few atomic
layers \cite{feder}. In this method one resolves
the quantum states of the incoming and outgoing electrons to extract
the accessible information on the sample under investigation.
On the other hand, a promising technique emerged in recent years where
an  electron pair, resolved in energy and momentum, is detected following
the impact of an 
 unpolarized electron beam upon a non-magnetic sample
\cite{wei,vos,stef,kir}.
As demonstrated successfully for a variety of materials \cite{wei,vos,stef},
the electron pair carries, under favorable conditions, direct signature of the
Bloch spectral function which is a central quantity as far as the electronic
structure is concerned.
However, these studies \cite{wei,vos} have been performed at higher energies
$(\approx 20\, keV)$ and 
the role of the spin polarization has not  been yet addressed.
Very recently, however,
 it has been demonstrated by a pioneering experiment \cite{sam99} that 
the electron-pair emission depends  strongly on
  the spin polarization of the
electron beam and the
magnetization of sample. 
Thus, it seems timely  to inspect theoretically the low-energy 
polarized electrons emission
from  ferromagnets.
We conclude: a) the electrons' spectra are  quantified fully  by 
a set of irreducible tensorial components; b)  under certain
circumstances specified below, the electron-pair spectrum is  directly
related to the spin-resolved spectral function of the surface. 

For a theoretical formulation 
we consider  a reaction in which two electrons are simultaneously
emitted from a ferromagnet with a defined magnetization direction
$\hat {\bf M}$  after the impact of a mono-energetic
spin polarized electron beam.  
The spins of the electrons in the incoming beam and in the sample
are assumed to be good quantum numbers.
A  corresponding experiment  resolves the
asymptotic  wave vectors of the impinging and the
two emitted (vacuum) electrons which we
 label  ${\bf k}_1$ and ${\bf k}_1'$, ${\bf k}_2'$, respectively.
No spin analysis of the outgoing electrons is performed.
The target surface is described by the
 state vectors $|\phi_{\epsilon,\alpha,s_2,m_{s_2}}\rangle$, where
$\epsilon$ is the ground state energy, 
 $m_{s_2}$ is the  projection  of the  spin ${\bf s}_2$ of the ground state
along a quantization axis and
$\alpha$ denotes all other quantum numbers.
The spin polarization of the incoming beam   with projection $m_{s_1}$
of the electron's spin ${\bf s}_1$
is characterized by the density matrix
$\rho^{s_1}_{m_{s_1}m_{s_1}}$ whereas the population of the 
magnetic sublevels of the 
state $|\phi_{\epsilon,\alpha,s_2,m_{s_2}}\rangle$ is
given by the density matrix $\bar\rho^{s_2}_{m_{s_2}m_{s_2}}$.
The scattering probability 
is  related to  (atomic units, a.u., are used throughout)
\begin{eqnarray}
W({\bf k}_1',{\bf k}_2'; {\bf k}_1)=
C \sum_{
m_{s_1'}, m_{s_2'},
m_{s_1}, m_{s_2}} \int\!\!\!\!\!\!\! { \sum_\alpha}\!\! &&
 {\cal M}({\bf k}_1',
{\bf k}_2',m_{s_1'}, m_{s_2'};\alpha,m_{s_2},
{\bf k}_1,m_{s_1})\nonumber\\
 &&\rho^{s_1}_{m_{s_1}m_{s_1}}\bar\rho^{s_2}_{m_{s_2}m_{s_2}}(\epsilon,\alpha)
 \nonumber\\
&&{\cal M}^*(
{\bf k}_1',
{\bf k}_2',m_{s_1'}, m_{s_2'};\alpha,m_{s_2},
{\bf k}_1,m_{s_1})\, \delta(E_f-E_i) 
\label{tdcs}\end{eqnarray}
where $ E_f$ is the final-state total energy. The initial-state total energy $E_i$ is
$E_i= E_{{\bf k}_1}-\epsilon $ whilst 
 $E_{{\bf k}_1}$ is the energy of the projectile
 beam and $C=(2\pi)^4 / k_1$. All energies are measured with respect
to the vacuum level.
  The transition amplitude 
 ${\cal M}({\bf k}_1',
{\bf k}_2',m_{s_1'}, m_{s_2'};\alpha,m_{s_2},
{\bf k}_1,m_{s_1})$ is given by
${\cal M}=\langle \psi_{{\bf k}_1',{\bf k}_2',m_{s_1'}, m_{s_2'}}
 | {\cal T} | \phi_{\epsilon,\alpha,s_2,m_{s_2}}\, \varphi_{{\bf k}_1,s_1m_{s_1}}\, 
 \rangle$ where
$\varphi_{{\bf k}_1,s_1m_{s_1}}$ is a spinor vacuum state describing the incoming beam.
The  emitted electrons with spin projections $m_{s_1'},m_{s_2'}$
are represented by the state vector $ |\psi_{{\bf k}_1',
{\bf k}_2',m_{s_1'},m_{s_2'}}\rangle$
whereas ${\cal T}$ is the transition operator.
 In Eq.(\ref{tdcs})
the density matrices are diagonal. This is not a restriction as they can always be diagonalized
by a rotation in the appropriate spin space. Furthermore, we adopt
$\hat {\bf M}$ as a joint quantization axis  for ${\bf s}_1$ and ${\bf s}_2$. 
 In case ${\bf s}_1$ and ${\bf s}_2$
do not have a common quantization axis, we apply an appropriate  spin
rotation of the density matrix  of the incoming beam.
For convenience we express  
the  electrons' final state  in  the total spin space as
$ \left.\left| \psi_{{\bf k}_1',{\bf k}_2',m_{s_1'}, m_{s_2'}}\right.\right\rangle
=\sum_{SM_S}\left.\left.\left\langle SM_S
 \right|   s_1'm_{s_1}',s_1'm_{s_1}'  \right   \rangle \left|
 \Psi_{{\bf k_1}',{\bf k_2}';SM_S}\right\rangle\right.  $
where $S$ is the total spin and $M_S$ is its projection.

To disentangle geometrical from dynamical 
properties we expand the density matrices
in state multipoles (statistical tensors) $\rho_{pq}$ \cite{fan},
\begin{eqnarray}
\rho^{s_1}_{m_{s_1}m_{s_1}} &=& \sum^{2s_1}_{p_1=0} (-)^{p_1-s_1-m_{s_1}} \langle s_1 -m_{s_1};
 s_1 m_{s_1} | p_1 q_1=0 \rangle
\rho_{p_1q_1=0}.\label{dens_1}\\
\bar\rho^{s_2}_{m_{s_2}m_{s_2}} (\epsilon,\alpha)\!\!\!&=&\!\!\! \sum^{2s_2}_{p_2=0} (-)^{p_2-s_2-m_{s_2}} \langle s_2 -m_{s_2};
 s_2 m_{s_2} | p_2 q_2=0 \rangle
\bar\rho_{p_2q_2=0}(\epsilon,\alpha).\label{dens_2}
\end{eqnarray}
Substituting
Eqs.(\ref{dens_1},\ref{dens_2})
into the general expression (\ref{tdcs})
yields 
\begin{equation}
W =
 \int\!\!\!\!\!\!\! { \sum_\alpha}\sum_{p_1=0}^{2s_1} \sum_{p_2=0}^{2s_2}\rho_{p_1q_1=0}\bar\rho_{p_2q_2=0}(\epsilon,\alpha)
 \Lambda^{p_1,p_2}_{q_1=0,q_2=0}\delta(E_f-E_i)
\label{sigpar}\end{equation}
where
\begin{eqnarray}
 \Lambda^{p_1,p_2}_{q_1=0,q_2=0}\!&=& 
 C\sum_{m_{s_1}}  (-)^{p_1-s_1-m_{s_1}} \left.\left\langle s_1 -m_{s_1};
 s_1 m_{s_1}\right| p_1 q_1=0 \right\rangle \nonumber\\
&&  \sum_{m_{s_2}}
 (-)^{p_2-s_2-m_{s_2}} \left.\left\langle s_2 -m_{s_2};
 s_2 m_{s_2} \right| p_2 q_2=0\right\rangle \nonumber\\
&& \!\!\sum_{SM_S}\!\!{\cal M}({\bf k}_1',{\bf k}'_2,
 SM_s;\alpha,m_{s_2}, {\bf k}_1, m_{s_1})
{\cal M}^*({\bf k}_1',{\bf k}'_2, SM_s;\alpha,m_{s_2}, {\bf k}_1, m_{s_1}).
\nonumber\\
\label{paramt2}
\end{eqnarray}
The decisive point is that 
the  sum over $m_{s_1}$ ($m_{s_2}$) in  Eq.(\ref{paramt2}) defines
 the component (along  $\hat {\bf M}$) of a  spherical tensor
of rank $p_1$  ($p_2$) \cite{berunpub}.  
This mathematical observation yields important information
as to the transformation  behaviour of $\Lambda^{p_1,p_2}_{0,0}$:
$\Lambda^{p_1=0,p_2}_{0,0}$ ($\Lambda^{p_1,p_2=0}_{0,0}$)
 is a {\em scalar} with respect to spin rotations
generated by ${\bf s}_1$ (${\bf s}_2$), i.e. it represents spin  averaged
quantities in the ${\bf s}_1$ (${\bf s}_2$) spin space,
 whereas the components $\Lambda^{p_1=odd,p_2}_{0,0}$ ($\Lambda^{p_1,p_2=odd}_{0,0}$)
can be regarded as a spin {\em orientation} in the
 ${\bf s}_1$ (${\bf s}_2$) spin space  (for $p_1=1$ 
it is a polar vector) and hence changes sign
upon spin reflection,
 i.e. $\Lambda^{p_1=odd,p_2}_{0,0}(-m_{s_1})=-
\Lambda^{p_1=odd,p_2}_{0,0}(m_{s_1})$ 
  [$\Lambda^{p_1,p_2=odd}_{0,q_0}(-m_{s_2})=-
\Lambda^{p_1,p_2=odd}_{0,0}(m_{s_2})$]. The tensorial
components with even $p_1$ values are alignment parameters, i.e.
they describe the deviations in the spectra from the unpolarized case. 
The above formalism is easily generalized \cite{berunpub} to the  
case of strong spin-orbit coupling   and/or multi-electron emission.
For two electrons
Eq.(\ref{sigpar}) reduces to
\begin{eqnarray}
W= \int\!\!\!\!\!\!\! { \sum_\alpha}\left\{\Lambda^{0,0}_{0,0}
\left[ \rho_{00}\, \bar\rho_{00} +
\rho_{00}\, \bar\rho_{10}
\, \frac{\Lambda^{0,1}_{0,0}}{\Lambda^{0,0}_{0,0}}
 + \rho_{10}\bar\rho_{00}
 \frac{\Lambda^{1,0}_{0,0}}{\Lambda^{0,0}_{0,0}}
+ \rho_{10}\bar\rho_{10}
\frac{\Lambda^{1,1}_{0,0}}{\Lambda^{0,0}_{0,0}}
\right]\delta(E_f-E_i )\right\} .
\label{norm}\end{eqnarray}
The first term of the sum in Eq.(\ref{norm}) is the pair emission rate averaged
over the spin orientation of the
incoming electron beam and the spin polarization of the
sample.
The second term describes the spin asymmetry  due to the 
inversion of the magnetization while the incoming electron
beam is being  {\em unpolarized }.
The third term
is the spin asymmetry in the
electron-pair emission from {\em unpolarized targets}
when inverting the spin polarization of the
electron beam.
 In absence of explicit spin 
interactions in the {\em transition operator} ${\cal T}$, e.g. spin-orbit coupling, 
the
parameters $\Lambda^{1,0}_{0,0}$ and $\Lambda^{0,1}_{0,0}$ 
vanish.
Finally the 
last term of  Eq.(\ref{norm}), the prime focus of the
following  calculations, is related to
the electron-pair emission from spin-polarized samples by
spin polarized electrons. It is a polar vector both in the
${\bf s}_1$ and ${\bf s_2}$ spin spaces, i.e.
$ \Lambda^{1,1}_{0,0}(-m_{s_1},m_{s_2})=
-\Lambda^{1,1}_{0,0}(m_{s_1},m_{s_2})=\Lambda^{1,1}_{0,0}(m_{s_1},-m_{s_2})$.
The explicit forms of $\Lambda^{1,1}_{0,0}$ and 
$\Lambda^{0,0}_{0,0}$
are derived from Eq.(\ref{paramt2}) to be
\begin{eqnarray}
\Lambda^{1,1}_{0,0}&=& \! \frac{C}{2}\sum_{S=0}^1\sum_{M_s}\left\{
|{\cal M}({\bf k}_1',{\bf k}_2',SM_s;{\bf k}_1,\alpha,\downarrow,\Downarrow)|^2 - |{\cal M}({\bf k}_1',{\bf k}_2',SM_s;{\bf 
k}_1,\alpha,\uparrow,\Downarrow)|^2\right.\nonumber\\
&& \left.+|{\cal M}({\bf k}_1',{\bf k}_2',SM_s;{\bf k}_1,\alpha,
\uparrow,\Uparrow)|^2 -|{\cal M}({\bf k}_1',{\bf k}_2',SM_s;{\bf k}_1,\alpha,\downarrow,\Uparrow)|^2\right\}
\label{lam}\\
\Lambda^{0,0}_{0,0}&=& \! \frac{C}{2}\sum_{S=0}^1\sum_{M_s}\left\{
|{\cal M}({\bf k}_1',{\bf k}_2',SM_s;{\bf k}_1,\alpha,\downarrow,\Downarrow)
|^2 +|{\cal M}({\bf k}_1',{\bf k}_2',SM_s;{\bf k}_1,\alpha,\uparrow,\Downarrow)|^2\right.\nonumber\\
&& \left. +|{\cal M}({\bf k}_1',{\bf k}_2',SM_s;{\bf k}_1,\alpha,
\uparrow,\Uparrow)|^2 +|{\cal M}({\bf k}_1',{\bf k}_2',SM_s;
{\bf k}_1,\alpha,\downarrow,\Uparrow)|^2\right\}
.\label{sum}\end{eqnarray}
 The  projections of the spins of the
sample state and the electron beam parallel (anitparallel) to the quantization axis
are labeled, respectively by the arrows $\Uparrow$ ($\Downarrow$) and $\uparrow$ ($\downarrow$).
In the total spin space
Eqs.(\ref{lam},\ref{sum}) are expressed in terms of the
singlet and the triplet partial
 cross sections, $X^{(S=0)}$ and $X^{(S=1)}$, respectively, i.e.
\begin{eqnarray}
\Lambda^{1,1}_{0,0}&=& \frac{C}{2}\left[X^{(S=1)}({\bf k}_1',{\bf k}_2';{\bf k}_1;\alpha)-X^{(S=0)}({\bf k}_1',{\bf k}_2';{\bf 
k}_1;\alpha)\right]
\label{lamc}\\
\Lambda^{0,0}_{0,0}&=& 
\frac{C}{2}\left[3X^{(S=1)}({\bf k}_1',{\bf k}_2';{\bf k}_1;
\alpha)+X^{(S=0)}({\bf k}_1',{\bf k}_2';{\bf k}_1;\alpha)\right]=: 2X^{{\it tot}}
.\label{sumc}\end{eqnarray}
$X^{(S=0)}$ and $X^{(S=1)}$ are determined by the matrix elements, 
$T^{(S)}({\bf k}_1',{\bf k}_2';{\bf k}_1,\alpha)$, of the
 singlet ($S=0$) and triplet ($S=1$)  transition operators  
${\cal T}^S= ({\bf 1} + (-1)^S{\cal P}_{12}) {\cal T}$. Here  
${\cal P}_{12}$ is a permutation operator that interchanges the two emitted electrons.
Thus, one obtains the symmetry property
$ T^{(S)}({\bf k}_1',{\bf k}_2';{\bf k}_1;\alpha)
= (-1)^S 
T^{(S)}({\bf k}_2',{\bf k}_1';{\bf k}_1,\alpha)$, i.e. in situations where an interchange
of the electrons does not modify the ionization dynamics the triplet scattering amplitude
and hence $X^{(S=1)}=C|T^{(S=1)}|^2$ vanishes.
An example will be shown below. 
Till this point the electronic and structural properties of the sample have not been  yet 
specified. For perfect clean surfaces the integral over $\alpha$ in
Eq.(\ref{norm}) implies summation over the
surface Bloch   wave vector ${\bf k}_{2\parallel}$ and over the surface layers.
 The Bloch theorem imposes a conservation law
for the surface components of the
{\em total} wave vector of the emitted electrons ${\bf K}^+_\parallel=
{\bf k}_{1\parallel}'+{\bf k}_{2\parallel}'$ \cite{diff98}, i.e.
the change of ${\bf K}^+_\parallel$ from its 
initial value ${\bf k}_{1\parallel}+{\bf k}_{2\parallel}$ (before the collision)
 is restricted to  a multiple of the surface reciprocal lattice vector
${\bf g}_\parallel$. This fact can be used to perform the integrals over ${\bf k}_{2\parallel}$
in Eq.(\ref{norm}). 
Therefore, in absence of spin interactions in the Operator ${\cal T}$
 Eq.(\ref{norm}) reduces to a summation over the surface layers, indexed by $l$, and over 
$g_{\parallel}$, i.e.
\begin{eqnarray}
W\!\! \propto \!\!\!  \sum_{{\bf g}_{\parallel},l}
\left\{2X^{{\it tot}}({\bf k}_1',{\bf k}_2';{\bf k}_1, {\bf g}_{\parallel},l)
\left[ \rho_{00}\, \bar\rho_{00}(\epsilon,{\bf \Lambda}_\parallel,l) +
\rho_{10}\bar\rho_{10}(\epsilon,{\bf \Lambda}_\parallel,l)
 A^s({\bf k}_1',{\bf k}_2';{\bf k}_1, {\bf g}_{\parallel},l)
\right]\delta(E_f-E_i )\right\} 
\label{as}\end{eqnarray}
where ${\bf \Lambda}_\parallel={\bf K}^+_\parallel -{\bf g}_\parallel-{\bf k}_{1\parallel}$. 
 The "exchange scattering asymmetry"  has been defined as 
\begin{equation}
A^s:=\frac{
X^{(S=1)}({\bf k}_1',{\bf k}_2';{\bf k}_1, {\bf g}_\parallel
,l)-X^{(S=0)}(
{\bf k}_1',{\bf k}_2';{\bf k}_1, {\bf g}_{\parallel},l)}{ 
 3X^{(S=1)}({\bf k}_1',{\bf k}_2';{\bf k}_1, {\bf g}_{\parallel},l)+X^{(S=0)}(
{\bf k}_1',{\bf k}_2';{\bf k}_1, {\bf g}_{\parallel},l)}.
\label{asex}\end{equation}

To calculate the terms in Eq.(\ref{as}) the 
state multipoles $\rho_{10}$ and $\bar\rho_{10}$ are needed. 
These can be obtained by inverting
the relations (\ref{dens_1},\ref{dens_2}). 
 In the standard representation, the density operators of the beam  and the surface 
are linearly expanded in
 terms of the Pauli matrices $ {\bf \sigma}$ 
as $\rho^{s_1}={\bf 1} + {\bf P}_1\cdot{\bf \sigma}$
and $\bar\rho^{s_2}=
w_0({\bf k}_{2\parallel},l,\epsilon)
({\bf 1} + {\bf P}_2\cdot{\bf \sigma})$ where $
w_0({\bf k}_{2\parallel},l,\epsilon)$ is the   spin-averaged
Bloch spectral function of the layer $l$ and ${\bf P}_1$ and ${\bf P}_2$ are the
polarization vectors.
 The  sample polarization  is given by
$P_2=[w({\bf k}_{2\parallel},l,\epsilon,\Uparrow)
-w({\bf k}_{2\parallel},l,\epsilon,\Downarrow)]
/[w_0({\bf k}_{2\parallel},l,\epsilon)]$. 
Here $w({\bf k}_{2\parallel},l,\epsilon,m_{s_2})$
stands for the spin and layer resolved Bloch spectral function. Thus we obtain
$\rho_{00}\, \bar\rho_{00}=[w_0({\bf k}_{2\parallel},l,\epsilon)]/2$
and $\rho_{10}\bar\rho_{10}=[w_0({\bf k}_{2\parallel},l,\epsilon)]\,  P_1 P_2/2$ and
Eq.(\ref{as}) reduces to 
\begin{eqnarray}
W \propto \sum_{{\bf g}_{\parallel},l} \, w_0({\bf \Lambda}_\parallel,l,\epsilon)\, 
X^{{\it tot}}[ 1  +{\cal A} ]
\delta(E_f-E_i ) .
\label{totfin}\end{eqnarray}
The asymmetry function ${\cal A}$ has been introduced as
\begin{equation}
{\cal A} = P_1\frac{\sum_l \left[ w({\bf \Lambda}_\parallel,l,\epsilon,\Uparrow)
-w({\bf \Lambda}_\parallel,l,\epsilon,\Downarrow)\right] \sum_{{\bf g}_{\parallel}}
X^{{\it tot}} A^s \delta(E_f-E_i )}
{\sum_{l'}w_0({\bf \Lambda}_\parallel,l';,\epsilon)\, \sum_{{\bf g}_{\parallel}'}
 X^{{\it tot}} \delta(E_f-E_i )}
=\frac{W(\uparrow\Uparrow)-W(\downarrow\Uparrow)}{
W(\uparrow\Uparrow)+W(\downarrow\Uparrow)}
.\label{asfin}\end{equation}
Thus, for the calculation of the  tensorial parameters 
 two major ingredients are needed: 1) The spin and layer-resolved
  spectral function of the sample which 
can be obtained from the trace of the imaginary part of the
corresponding Green function and 2) the matrix element of the singlet and triplet
transition operators.  Now we calculate the terms in Eq.(\ref{totfin})
for  a Fe(110) surface.  The Bloch spectral functions
used here are provided by two independent calculations:
1) The scalar relativistic full-potential linearized augmented plane-wave method (FPLAPW) 
\cite{hubner,bla}
and 2) the  full relativistic layer Korringa-Kohn-Rostoker method (LKKR) \cite{feder,tam92}. 
For the calculations of the transition matrix element we approximate the ${\cal T}$ operator
by ${\cal T} = U_{\it surf} + U_{ee} ({\bf 1} + G^-_{ee}U_{\it surf})$ where
$U_{ee}$ is the electron-electron interaction, $  G^-_{ee}$ is the Green function of the
electron pair and $U_{\it surf}$ is the surface scattering potential. For $U_{\it surf}$ we 
employ, for a given layer,  a
linear combination of non-overlapping  muffin-tin potentials \cite{diff98}.

As stated above, for certain geometries,
  the triplet scattering amplitude vanishes 
due  to symmetry 
and hence $A^{s}$ 
tends to $-1$ (cf. Eq.(\ref{asex})). Thus,
 if a  monolayer or a bulk system
is  considered the magnetic asymmetry $P_2$
 can be scanned by determining 
$W(\uparrow\Uparrow)$ and $W(\downarrow\Uparrow)$.  This yields
a direct (relative) estimate of the
population of the spin states in the sample. For multilayered systems,
we have to consider the weighting factor $X^{{\it tot}}$
in Eq.(\ref{asfin}). 
An example is shown in Fig.1 for a Fe(110) sample. The two electrons
are detected with fixed {\em equal} energies
in the $x-z$ plane and at {\em symmetric} positions with respect to the $z$ direction
while the incident beam direction is varied in the $z-y$ plane. The
 experiment, in the geometry of   Fig.1,
is invariant under  a $180^{\rm o}$ rotation with respect to the $z$ direction.
This rotation can be regarded as  an interchange of ${\bf k}_1'$ by ${\bf k}_2'$ and since
$T^{(S=1)}({\bf k}_1',{\bf k}_2';{\bf k}_1,\alpha) =-
T^{(S=1)}({\bf k}_2',{\bf k}_1';{\bf k}_1,\alpha)$ the triplet scattering ($X^{(S=1)}=
C|T^{(S=1)}|^2$) vanishes.

The energies $\epsilon$ in
Eq.(\ref{asfin}) is determined by $\epsilon=E_1'+E_2'-E_{{\bf k}_1}$ where $E_1'$ and 
$E_2'$ are the energies of the vacuum electrons. Thus we tune
$E_1'$, $E_2'$ and $E_{{\bf k}_1}$ such that $\epsilon$ coincides with the Fermi energy $E_F$.
Now by varying $\beta=\cos^{-1}\hat{\bf z}\cdot\hat{\bf k}_1$ we 
scan  $P_2$  along the
$\Gamma$-$N$ direction in the  Brillouin zone, as shown in Fig.1.
Alternatively one may fix the direction ${\bf \Lambda}_{\parallel}
={\bf k}_{1\parallel}$
and image $ P_2(\epsilon)$ by varying, e.g., the incident energy.
 For a polarized homogeneous electron gas
one scans (as function of energy) 
 the relative difference between the occupied density of states
of the majority and minority bands.

Away from the points of high symmetry (cf. Fig.1)
 the scattering dynamics, as described
by  $X^{(S=0)}$ and $X^{(S=1)}$ become dominant. 
An example is shown in Fig.2 for $\beta=0$.
Again at the $\Gamma$ point 
(${\bf k}_{1\parallel}'=-{\bf k}_{2\parallel}'$) the asymmetry ${\cal A}$,
and in particular its sign,
is determined solely by $P_2$. The 
For highly asymmetric 
energy sharing the scattering exchange asymmetry $A^s$ is small which leads to a
reduced asymmetry ${\cal A}$, as seen in Fig.2.

I would like to thank X. Qian, W. H\"ubner, A. Ernst and N. Fominykh for
 communicating their results on the spectral functions and them
and S. N. Samarin, J. Henk, J. Kirschner, and P. Bruno for valuable discussions.

% \end{multicols}
%                        Figure Captions
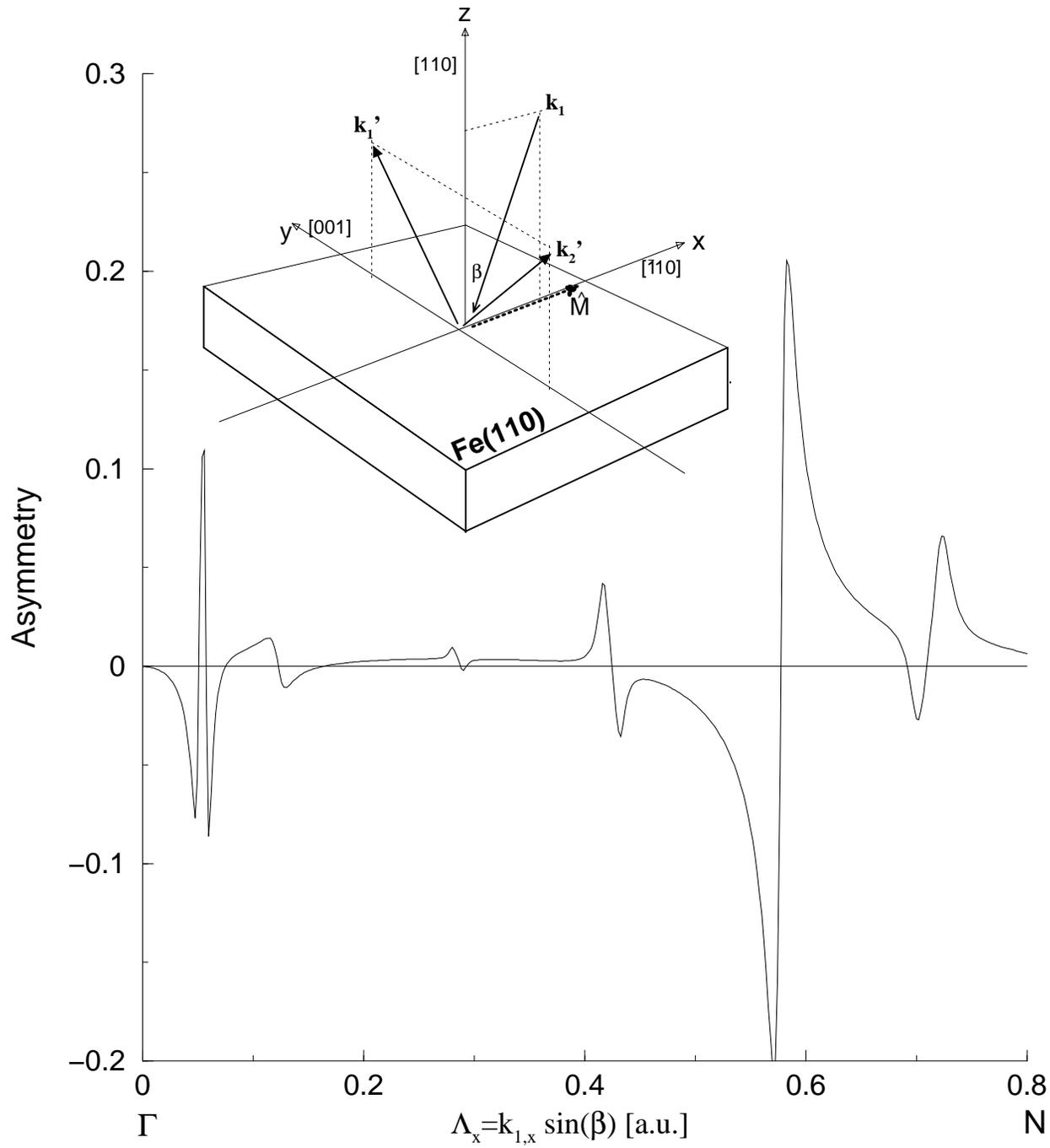
\noindent{\bf Fig.\ 1:}
%\figure{ }
The asymmetry, as given by Eq.(\ref{asfin}),
  for the emission of two equal-energy electrons from a magnetized
Fe(110) surface following the
impact of a polarized electron beam with an  energy
of $35$ eV. The total energy of the pair is fixed to
 $E_1'+E_2'=25.15$ eV. The two electrons are detected in the
$y$-$z$ plane at  symmetric
position $\cos^{-1}\hat{\bf z}\cdot\hat{\bf k}_1'=40^{\rm o}=
\cos^{-1}\hat{\bf z}\cdot\hat{\bf k}_1'$ 
left and right to the $z$ axis (cf. inset) and 
$\hat {\bf M}\parallel x$.
 The angle of incidence   $\beta=\cos^{-1}
\hat{\bf z}\cdot\hat{\bf k}_1$ is varied in the $x-z$ plane, as shown
by the inset. In this geometry, the triplet scattering vanishes and ${\cal A}$
can be related to $P_2$.
The  predominant contributions to ${\cal A}$ originate from the 
first and second surface layers.

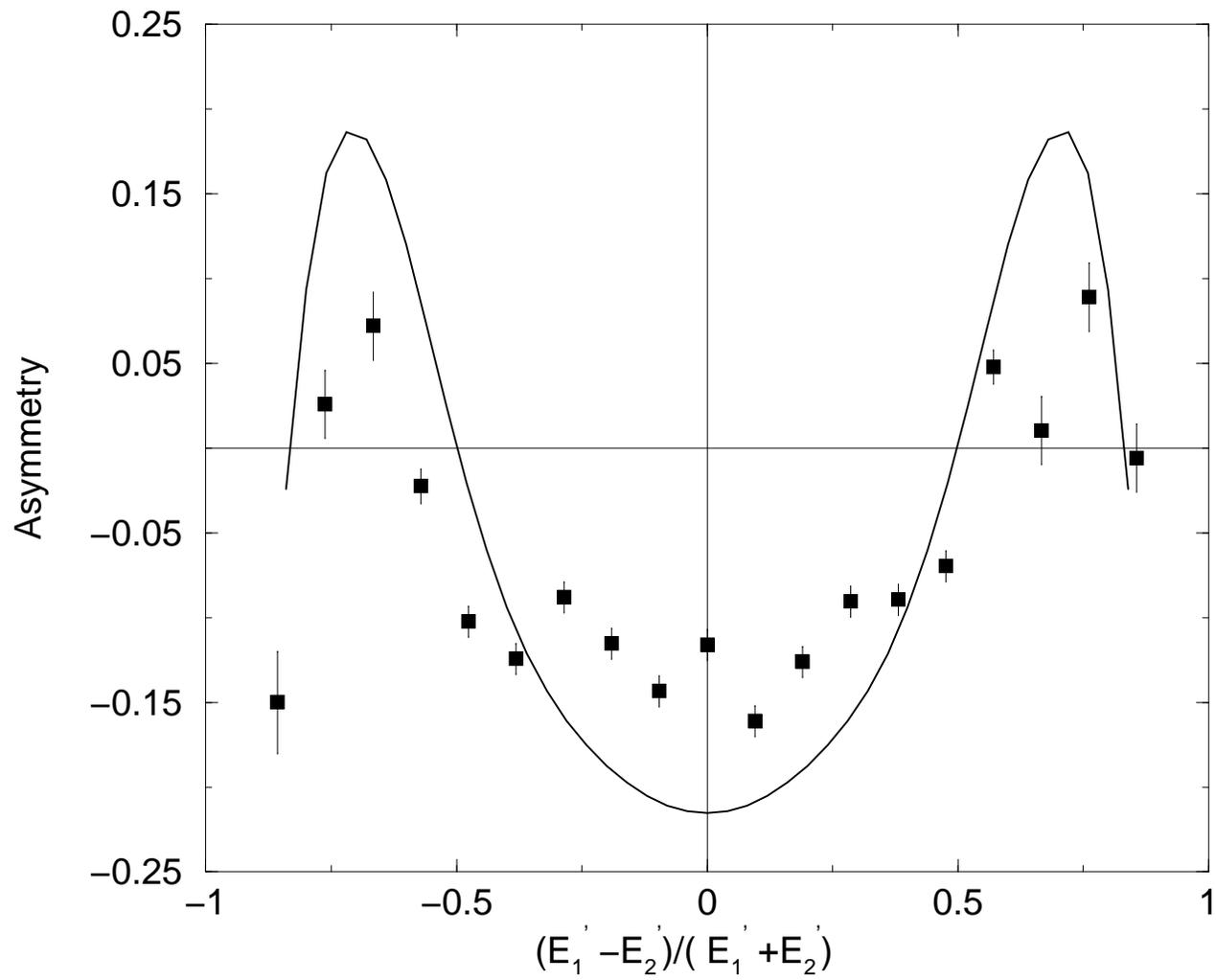
\noindent{\bf Fig.\ 2:}
%\figure{ }
%
The spin asymmetry ${\cal A}$ as function of the energy sharing
$(E_1'-E_2')/(E_1'+E_2')$ for a fixed total energy 
$(E_1'+E_2')=21$ eV. The incident electron  has an energy  
26 eV and a polarization degree of $\approx 65\%$.
The same target as in Fig.1, however, we choose  $\beta =0$ and
the two electron detectors to lay in the $x-z$ plane. As in Fig.1, 
the detectors are positioned at
$\cos^{-1}\hat{\bf z}\cdot\hat{\bf k}_1'=40^{\rm o}=\cos^{-1}\hat{\bf z}\cdot\hat{\bf k}_1'$.
The theoretical results are averaged over the angular resolution of the
detectors.
The experimental data are courtesy of Ref.\cite{sam99}. The spectral functions
are calculated within the scalar relativistic FLAPW method \cite{hubner}.

%
% \end{multicols}

\begin{thebibliography}{99}
\bibitem{hubert} A. Hubert and R. Sch\"afer
{\em magnetic domains: The analysis of magnetic microstructures}
(Springer Verlag, Berlin, 1998)

\bibitem{feder} 
{\em Polarized Electrons in Surface Physics}, edited by
R. Feder (World Scientific, Singapore, 1985).
\bibitem{wei} I. E. McCarthy  and E. Weigold
              Rep. Prog. Phys. {\bf{54}}, 789 (1991)

\bibitem{vos} M. Vos and I. E. McCarthy,
              Rev. Mod. Phys. {\bf{67}}, 713 (1995)
\bibitem{stef} A.S. Kheifets, S. Iacobucci, A. Ruoccoa, R. Camilloni, and G.
Stefani, Phys. Rev. B {\bf 57}, 7380 (1998)
\bibitem{kir}J. Kirschner, O. M. Artamonov and S. N. Samarin, Phys. Rev. Lett.
 {\bf 75}, 2424 (1995)
\bibitem{sam99}S. N. Samarin and J. Kirschner {\em private communication}
\bibitem{fan} U. Fano, Rev. Mod. Phys. {\bf 29} , 76  (1957)
\bibitem{berunpub}J. Berakdar {\em unpublished}
\bibitem{diff98}J. Berakdar,  S. N. Samarin,
 R.  Herrmann,and J. Kirschner: Phys. Rev. Lett.  {\bf 81}, 3535 (1998) 
\bibitem{hubner} X. Qian and W. H\"ubner {\em private communication} and
Phys. Rev. B (to be published)
\bibitem{bla}  P. Blaha, K. Schwarz, P. Sorantin, and S. B. Trickey,
Comput. Phys. Commun. {\bf 59}, 399 (1990)

\bibitem{tam92} E. Tamura in {\it 
Applications of Multiple Scattering Theory to
                  Materials Science} pp. 347, Eds. W. H. Butler, 
P. H. Dederichs, A. Gonis and
                  R. L. Weaver (Materials Research Society,
                        Pittsburgh, Pennsylvania, 1992)
% \bibitem{streun} M. Streun, G. Baum, W. Blask, and J. Berakdar,
% Phys. Rev. A {\bf 59} R4109 (1999)





\end{thebibliography}
\end{document}